%% file: main.tex
\documentclass{article}

\usepackage{microtype}
\usepackage{graphicx}
\usepackage{subfigure}
\usepackage{booktabs} % for professional tables
\usepackage{glossaries}

\usepackage{hyperref}

\usepackage[accepted]{synsml2023}

\usepackage{amsmath}
\usepackage{amssymb}
\usepackage{mathtools}
\usepackage{amsthm}

\usepackage[capitalize,noabbrev]{cleveref}

\theoremstyle{plain}

\theoremstyle{definition}

\theoremstyle{remark}

\usepackage{comment}
\usepackage{todonotes}

\glsdisablehyper
\loadglsentries{glossary}

\synsmltitlerunning{Evaluating the diversity and utility of materials proposed by generative models}

\begin{document}

\twocolumn[
\synsmltitle{Evaluating the diversity and utility of materials proposed by generative models}

\synsmlsetsymbol{equal}{*}

\begin{synsmlauthorlist}
\synsmlauthor{Alexander New}{yyy}
\synsmlauthor{Michael J. Pekala}{yyy}
\synsmlauthor{Elizabeth A. Pogue}{yyy}
\synsmlauthor{Nam Q. Le}{yyy}
\synsmlauthor{Janna Domenico}{yyy}
\synsmlauthor{Christine D. Piatko}{yyy}
\synsmlauthor{Christopher D. Stiles}{yyy}

\end{synsmlauthorlist}

\synsmlaffiliation{yyy}{Research and Exploratory Development Department, Johns Hopkins University Applied Physics Laboratory, Laurel, Maryland, USA}

\synsmlcorrespondingauthor{Alexander New}{alex.new@jhuapl.edu}

\synsmlkeywords{Machine Learning}

\vskip 0.3in
]

% this must go after the closing bracket ] following \twocolumn[ ...

% This command actually creates the footnote in the first column
% listing the affiliations and the copyright notice.
% The command takes one argument, which is text to display at the start of the footnote.
% The \synsmlEqualContribution command is standard text for equal contribution.
% Remove it (just {}) if you do not need this facility.

\printAffiliationsAndNotice{}  % leave blank if no need to mention equal contribution
% \printAffiliationsAndNotice{\synsmlEqualContribution} % otherwise use the standard text.

% \alex{title is bad and should be made better, etc. etc}
% \christine{wondering about a word like utility vs usability in the title? or even as specific as stability? realizable? that would flow down to many other paragraphs though}
% \christine{can the intro of the paper be tweaked to be more explicit about the tie to the theme of the workshop "combining scientific and machine learning"? What should we point to as the "scientific model"  here? The database data generated by first-principles models?

% "The Synergy of Scientific and Machine Learning Modeling Workshop (“SynS \& ML”) is an interdisciplinary forum for researchers and practitioners interested in the challenges of combining scientific and machine-learning models. The goal of the workshop is to gather together machine learning researchers eager to include scientific models into their pipelines, domain experts working on augmenting their scientific models with machine learning, and researchers looking for opportunities to incorporate ML in widely-used scientific models."

% https://syns-ml.github.io/2023/cfp/}

\begin{abstract}
% This document provides a basic paper template and submission guidelines.
% Abstracts must be a single paragraph, ideally between 4--6 sentences long.
% Gross violations will trigger corrections at the camera-ready phase.
Generative \gls{ML} models can use data generated by scientific modeling to create large quantities of novel material structures. Here, we assess how one state-of-the-art generative model, the \gls{PGCGM}, can be used as part of the inverse design process. We show that the default \gls{PGCGM}'s input space is not smooth with respect to parameter variation, making material optimization difficult and limited. We also demonstrate that most generated structures are predicted to be thermodynamically unstable by a separate property-prediction model, partially due to out-of-domain data challenges. Our findings suggest how generative models might be improved to enable better inverse design.
\end{abstract}

\glsresetall

\section{Introduction}\label{sec:introduction}

% \alex{note the workshop call and details in the scratchwork~\cref{sec:scratchwork}}

% \alex{limited to four pages, not counting references and appendices (but appendices should not be load-bearing)}

Inverse design---the discovery of materials with targeted properties---remains an important task in materials science~\cite{wang2022inverse}. In recent years, progress has been made, spurred by the use of \gls{ML} and iterative design loops% (\cref{fig:iterative_design}
~\cite{Pogue2022closedloop,Goodall2022discovery,Stanev2018superconductors,Attia2020closedloopbatteries,Zhang2020BO,Baird2022discover,Wines2023inverse}, but further challenges remain. A key reason is the lack of high-quality experimental and computational data, compared to similar fields like chemistry~\cite{Xie2022cdvae}.

% \begin{figure}
%     \centering
%     \includegraphics[width=0.6\linewidth]{figs/loop.png}
%     \caption{The iterative design process for inverse design of materials}
%     \label{fig:iterative_design}
% \end{figure}

% \christine{Perhaps can continue to connect to scientific modeling theme here:
% Quantum mechanical modeling and experimental characterization have enabled the generation of scientific modeling databases. These include MP, OQMD etc.)
% }
Computational and experimental characterization have enabled the generation of scientific materials databases like \gls{MP}~\cite{Jain2013mp}, \gls{OQMD}~\cite{Saal2013OQMD,Kirklin2015OQMD}, and \gls{ICSD}~\cite{Belsky2002icsd}, which contain hundreds of thousands of material structures. However, these still sample only a small fraction of possible materials.
% \alex{would be useful to have concrete number or citation here} 
% \lisa{It is a little unclear here. Maybe write "Thus, to expand number of materials capable of being sampled, it is natural to use these databases as inputs to generative ML models like GAN and VAE. These generative models predict new materials and assess their suitability for design tasks.}
To enable the discovery of additional materials structures, these scientific databases are used as training data for generative \gls{ML} models like \glspl{GAN}~\cite{Goodfellow2014gan} and \glspl{VAE}~\cite{Kingma2013vae}. Via additional property-prediction models, generated materials can be assessed for suitability in design tasks. Identified materials can be studied in detail, synthesized, and characterized. In recent years, a number of different \gls{ML} models for material generation have been developed~\cite{Zhao2021cubicgan,Kim2022scgan,Ren2022ftcp,Court2020conddfcvae,Zhao2023pgcgm,Alverson2023discovery}, and \gls{ML} has been successfully used for property prediction~\cite{Goodall2020Roost,New2022curvatureinformed,Xie2018CGCNN,Sanyal2018MTCGCNN,Chen2019megnet,Park2020ICGCNN,Choudhary2021alignn}.

In this work, we evaluate a state-of-the-art generative model, the \gls{PGCGM}~\cite{Zhao2023pgcgm} used to generate diverse and usable material structures. In~\cref{sec:generative_models}, we detail the generation process. In~\cref{sec:smoothness}, we identify a potential issue with the use of \gls{PGCGM} for inverse optimization---namely, a lack of smoothness with respect to property and validity variation in the model's input space. In~\cref{sec:stability}, we use a second \gls{ML} model to assess the stability of generated materials and highlight a lack of diversity in structures. We conclude with recommendations for how to enhance generative models for use in inverse design of materials.

\section{Generating novel material structures}\label{sec:generative_models}

Our study uses a state-of-the-art generative model, \gls{PGCGM}~\cite{Zhao2023pgcgm}. The \gls{PGCGM} extends the Wasserstein \gls{GAN}~\cite{Arjovsky2017wasserstein} to ternary material generation. Given three constituent atoms and space group of a ternary material system, \gls{PGCGM} predicts possible atom coordinates and lattice parameters. Generated unit cells are saved into \gls{CIF} format and can be read by tools like \texttt{pymatgen}~\cite{Ong2013pymatgen} and \texttt{jarvis}~\cite{Choudhary2020jarvis} for use and analysis. \gls{PGCGM} was shown to perform well in terms of material validity, diversity, and property similarity to training data, compared to other models like CubicGAN~\cite{Zhao2021cubicgan} and the \gls{FTCP}~\cite{Ren2022ftcp} method. \gls{PGCGM} trains a generator $f_\theta$ that satisfies:
\begin{equation}
    B, p = f_\theta(z, E, s),
    \label{eq:generator}
\end{equation}
where the outputs are: $B = (b_0, b_1, b_2) \in \mathbb{R}^{3\times3}$, the atom fractional coordinates for each constituent element; and $p =(a,b,c,\alpha,\beta,\gamma)$, the structure's lattice parameters; and the inputs are: $z \in \mathbb{R}^{128}$, a vector with entries sampled from a standard normal distribution; $E$, the three constituent atoms; and $s$, a space group. Thus, $B$ and $p$ can be used to construct a generated material's \gls{CIF}.

We use the pretained \gls{PGCGM} model and post-processing scripts available on GitHub\footnote{\url{https://github.com/MilesZhao/PGCGM}} and generate two sets of material structures. In the first, we fix the constituent atoms and the space group and generate $5,000$ samples from $f_\theta$ for the $\mathrm{Ta{-}Ge{-}As}$ system with the $Pm\bar{3}m$ space group; of these, $241$ are valid material structures\footnote{Following the post-processing scripts of~\cite{Zhao2023pgcgm}, by ``valid'', we mean that their estimated space group (from \texttt{pymatgen}) matches their input space group $s$, and atoms of the same type that occupy approximately the same spatial location can be merged together.}. We analyze how the \gls{GAN} sample $z$ impacts material structure of these materials in~\cref{sec:smoothness}. In the second set, we sample sets of constituent atoms and space groups and generate $500,000$ samples from $f_\theta$; of these, $27,116$ are valid. The validity rate ($27,116/500,000 \approx 5.4\%$) for this sample is slightly lower but comparable to the $7.14\%$ validity rate~\cite{Zhao2023pgcgm} report.

We analyze the predicted stability of these structures in~\cref{sec:stability}. \Cref{fig:lizn2pt} and~\cref{fig:sin2o2} in~\cref{sec:supplemental_figures} show example structures generated by \gls{PGCGM}.

\section{Assessing smoothness of GAN input space}\label{sec:smoothness}

% \alex{minor mathematical sin: ``smoothness'' is a well-defined term that we're slightly overloading/misusing; open to suggestions}

In the materials design problem, we look for a material that maximizes a suitability or property function $\mathrm{Outcome}$ (e.g., critical temperature for superconductors or solubility for drug-like molecules). Using \gls{PGCGM}, this problem can be formulated as: 
\begin{equation}
    \hat{B},\hat{p} = \arg\max_{z,E,s} \mathrm{Outcome}(B(z, E, s), p(z, E, s)),
    \label{eq:design}
\end{equation}
where $B, P, z, S, E$ are defined in Eq.~\ref{eq:generator}. Given a differentiable suitability function $\mathrm{Outcome}$ and fixed space group $s$ and elements $E$, Eq.~\ref{eq:design} suggests the use of gradient-based methods to optimize $\mathrm{Outcome}$ by differentiating it with respect to $z$. Similar approaches are often used with \glspl{VAE} for inverse design~\cite{GomezBombarelli2018vae,Xie2022cdvae} and with \glspl{GAN} for image editing~\cite{Zhu2020editing}.

Effective optimization requires two conditions be met: (1) small perturbations to a \gls{PGCGM} input $z$ corresponding to a valid (invalid) \gls{CIF} should continue to produce a valid (invalid) \gls{CIF}, and (2) perturbing a \gls{PGCGM} input $z$ corresponding to a valid \gls{CIF} should not drastically change properties of the unit cell (e.g., the gradient of $\mathrm{Outcome}$ should be Lipschitz-continuous with respect to $z$, which can enable convergence of gradient descent techniques in combination with other regularity conditions~\cite{Carmon2020bounding,Arjevani2023globallipschitz}).
% \mike{you might be able to speak about this in terms of ``stability'' (or ``smoothness'') and perhaps even give it a definition (e.g. $|| p(z_i) - p(z_j)|| \leq c ||z_i - z_j||$ with some constraint on $c$ since this is a finite dimensional space ? or whatever you have in mind).  For (1) it will not hold everywhere but maybe there's another way to characterize that.}

\subsection{Global smoothness analysis}\label{sec:global}

\begin{figure}
    \centering
    \includegraphics[width=0.7\linewidth]{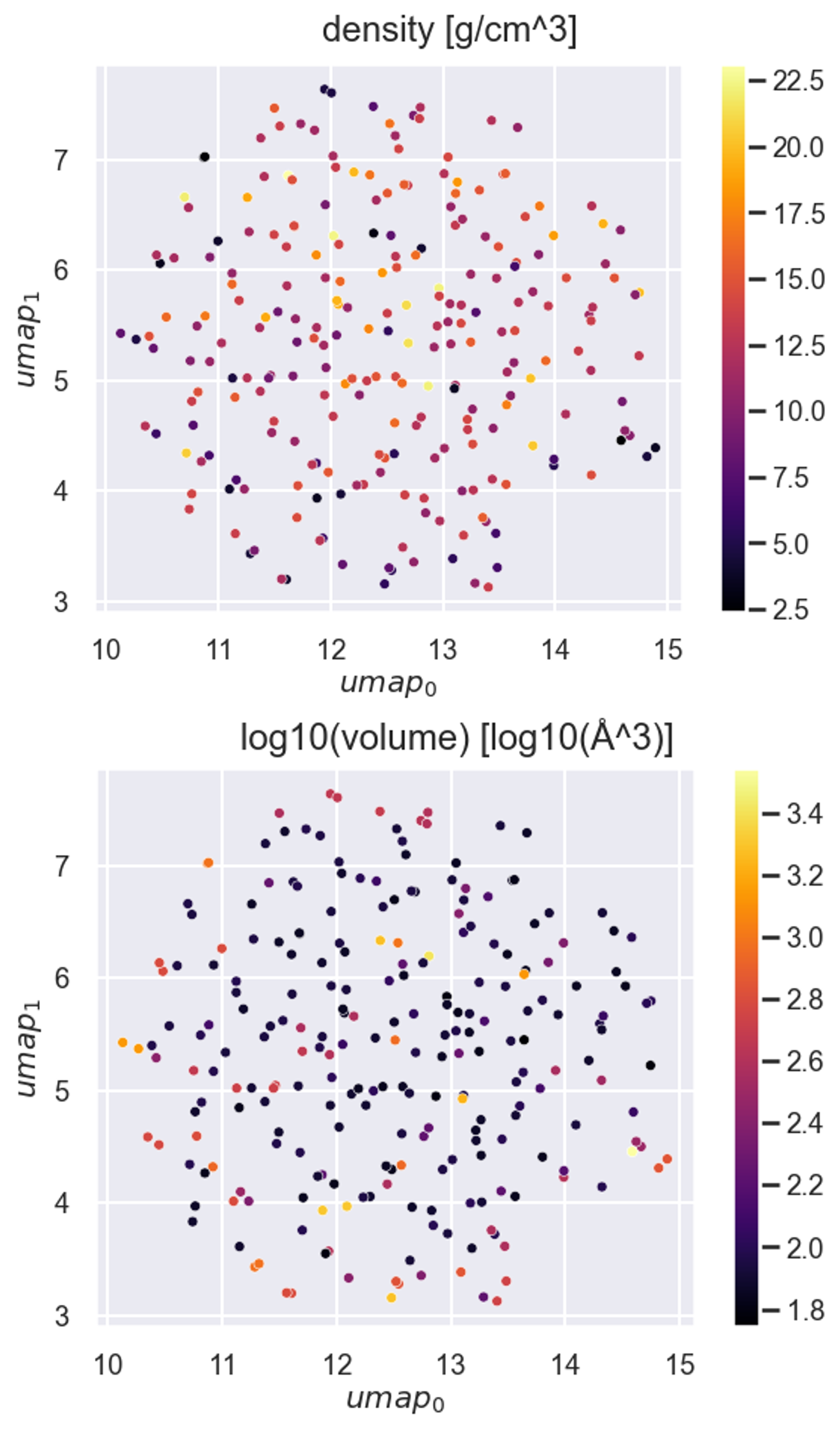}
    \caption{We use \gls{UMAP} to project \gls{PGCGM} inputs ($z$ in Eq.~\ref{eq:generator}) corresponding to valid \glspl{CIF} from the $\mathrm{Ta{-}Ge{-}As}$ system into two dimensions. Nearby points often have different property values, suggesting that the \gls{PGCGM} input-space is not generally smooth with respect to variation in material properties, making \gls{PGCGM} difficult to use as part of gradient-based material optimization. Statistical analyses (\cref{sec:smoothness}) further support this observation.}
    \label{fig:umap_properties}
\end{figure}

Using the 241 \gls{PGCGM} inputs $z$ corresponding to valid structures, we find that the pretrained \gls{PGCGM}'s input space does not appear to be globally smooth. In~\cref{fig:umap_properties}, we take the 241 \gls{PGCGM} inputs $z$ corresponding to valid structures and project them into two dimensions using \gls{UMAP}~\cite{Mcinnes2020umap}. It is common for a point and its nearest neighbor in the embedding to have very different property values.

We also conduct a more extensive analysis on  the relationship between the \gls{PGCGM} input space and the set of samples $z\in\mathbb{R}^{128}$ corresponding to valid materials. First, we use the \gls{KS} test to see if the distribution of each component of the samples $z$ was distinct from the standard normal distribution that generated all of the samples and adjust the $128$ $p$-values with the Benjamini-Hochberg~\cite{Benjamini1995fdr} scheme. Four $z$ components (out of $128$) were found to be significantly distinct from the standard normal distribution, suggesting that $z$ samples corresponding to valid structures and invalid structures are often statistically similar, which is evidence against the first condition. We also train random forest models to predict unit cell density and volume given $z$; the density-prediction model had an out-of-bag $R^2$ estimate of $0.16$, and the volume-prediction model had an out-of-bag $R^2$ estimate of $0.20$. This suggests that the relationship between \gls{PGCGM} input $z$ and resultant unit cell's properties is difficult to model, providing evidence against the second condition. We repeated this analysis with other material systems and saw comparable results.

\subsection{Local smoothness analysis}\label{sec:local}

Our conclusions in~\cref{sec:global} relied on a projection of a high-dimensional space to assess smoothness. To provide additional perspective, we perform a perturbation-based local smoothness analysis. This enables us to explore property variation in local neighborhoods of \gls{GAN} inputs $z$ corresponding to valid material structures. Specifically, we start by sampling a point $z_0$ from the \gls{GAN}'s input distribution $\mathcal{N}(0_{128}, I_{128})$. We sample $N$ random directions $r_n$ from the surface of the unit sphere in $\mathbb{R}^{128}$ and construct a uniform discretization $0<\varepsilon_1<\cdots<\varepsilon_M$ of the range $(0, \varepsilon_M]$. By evaluating the \gls{PGCGM} at each point $z_0 + \varepsilon_m r_n$ and calculating $\mathrm{Outcome}$ of the resultant material $f_\theta(z_0 + \epsilon_m r_n, E, s)$, we see $\mathrm{Outcome}$'s behavior across balls of radius $\epsilon_m$.

These calculations let us estimate Lipschitz constants for property functions by evaluating $|\mathrm{Outcome}(f_\theta(z_0 + \varepsilon_m r_n, E, s) - \mathrm{Outcome}(f_\theta(z_0 + \varepsilon_{m'} r_{n'}, E, s)|$ vs. $||\epsilon_m r_n - \varepsilon_{m'} r_{n'}||$. Although convergence rates for optimization involve Lipschitz constants of function gradients (i.e., $\nabla_z \mathrm{Outcome}(f_\theta(z_0 + \varepsilon_m r_n, E, s)$), these are harder to obtain for \gls{PGCGM}: the \gls{PGCGM}'s post-processing non-smoothly merges spatially-adjacent atoms of the same type together. Future work could estimate some of these Lipschitz constants using specially-developed techniques with provable guarantees~\cite{Jordan2020lipschitzrelu,Jordan2021lipschitzgenerative}.

\begin{figure}[h]
    \centering
    \includegraphics[width=1\linewidth]{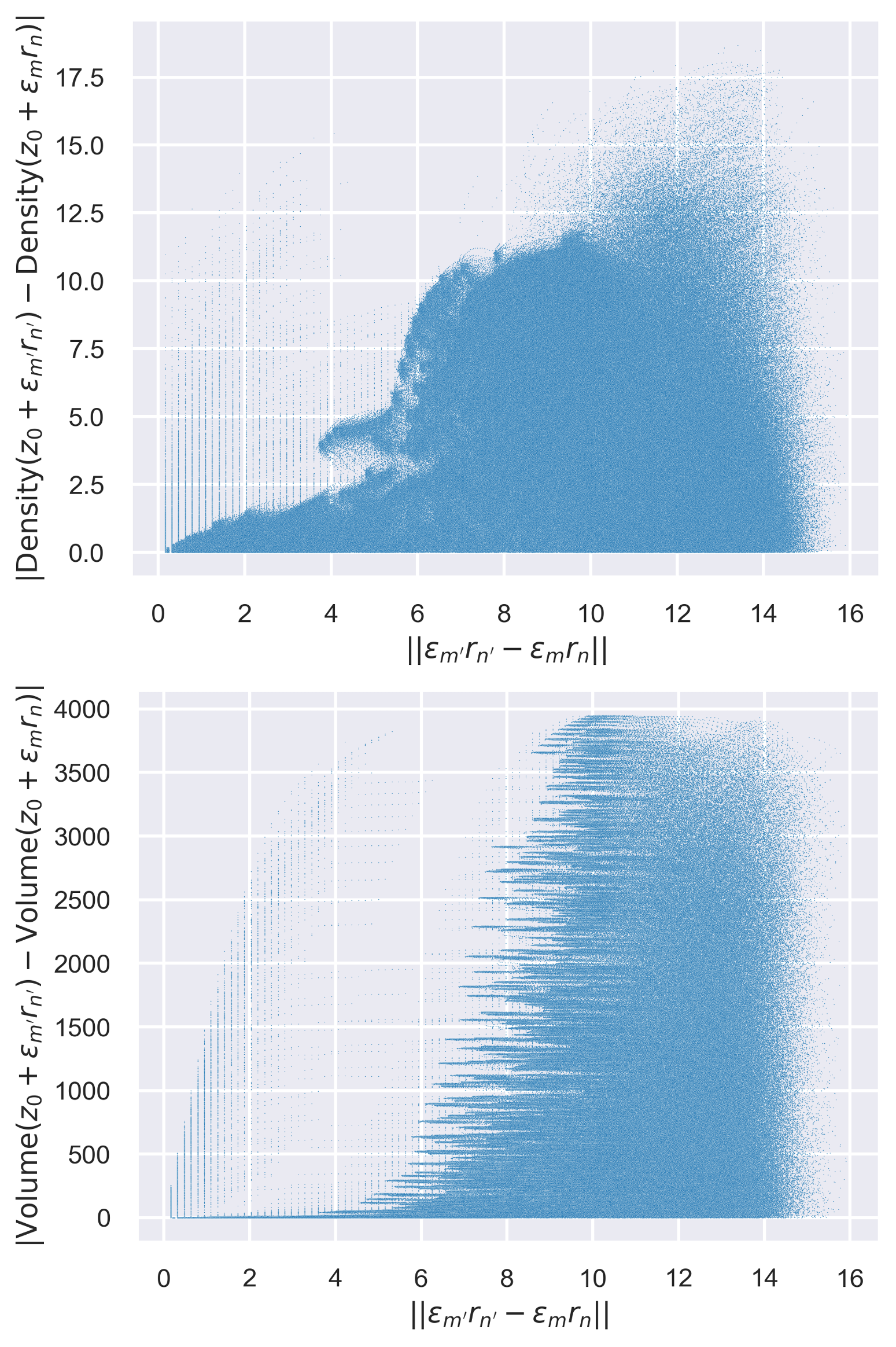}
    \caption{Using a sample from the $\mathrm{Ta{-}Ge{-}As}$ system, we analyze the local smoothness of the \gls{PGCGM}'s input space by plotting changes in material properties---density (top) and volume (bottom)---with respect to changes in the \gls{PGCGM}'s input $z$. Notably, the input space  features contours where $||\epsilon_m r_n - \varepsilon_{m'} r_{n'}||$ is constant, but the property changes.}
    \label{fig:pairwise_density}
\end{figure}

In this study, we select $z_0$ as one of the 241 stable materials sampled from the $\mathrm{Ta{-}Ge{-}As}$ system with the $Pm\bar{3}m$ space group used in~\cref{sec:global}, we sample $N = 128$ different directions $r_n$ and evaluate perturbations $r_n\epsilon_m$ up to a maximum radius of $\epsilon_M = 10$. In~\cref{fig:pairwise_density}, we show pairwise property and input distances for the resulting valid \glspl{CIF}. For smaller input distances ($||\epsilon_m r_n - \varepsilon_{m'} r_{n'}|| \lessapprox 4$), the \gls{PGCGM} input space is locally smooth in the sense that small changes in the input mostly lead to small changes in the property. After that, the range in variation of the properties increases significantly, which aligns with the global smoothness analysis of~\cref{sec:global}.

\section{High-throughput thermodynamic stability predictions}\label{sec:stability}

% \christine{reworded first sentence slightly - used to be "Generative models suggest too many structures to all ex- perimentally synthesize and characterize."}
Generative models suggest large numbers of potential material structures, and not all generated structures are guaranteed to be thermodynamically stable. Some of these can be further characterized by experimental or computational techniques, but this characterization is too expensive in time and cost to be performed for all or even most generated materials. Thus, we trained a \gls{ML} to predict the thermodynamic stability of the \gls{PGCGM}-generated structures.

% Experimentally characterizing large numbers of materials with computational or experimental techniques is expensive in time and cost. 
% Thus, generative models suggest many more structures than can be experimentally synthesized and characterized. First-principles calculations can be used to assess thermodynamical stability of novel structures, but doing this for all generated structures is also not computationally feasible.
% \alex{would be good to have more details here on scalability of first-principles} 
% Thus, we trained a stability-prediction \gls{ML} model to rapidly assess thermodynamic stability of generated structures.

In particular, we use \gls{ALIGNN}~\cite{Choudhary2021alignn} to predict decomposition enthalpy $\Delta H_d$. \Gls{ALIGNN} is trained on a set of ternary structures and computationally-predicted $\Delta H_d$ taken from \gls{MP} and collected by~\citealt{Bartel2020stability}. A structure is predicted to be thermodynamically stable if its decomposition enthalpy is negative.
% \alex{would be good to have more details on $\Delta H_d$ vs. other quantities like energy-under-convex hull}

The \gls{ALIGNN} forward pass is:
\begin{equation}
    \Delta H_d = \ell(h) = \ell(a(g)),
    \label{eq:alignn}
\end{equation}
where $\Delta H_d$ is the predicted decomposition enthalpy, $\ell$ is a linear layer, and $h = a(g) \in \mathbb{R}^{256}$ is the embedding of an input graph $g$ produced by edge-gated graph convolutional layers~\cite{Bresson2017edgegated}. Our \gls{ALIGNN} model is trained with the default architecture and hyperparameters.

\begin{figure}
    \centering
    \includegraphics[width=0.9\linewidth]{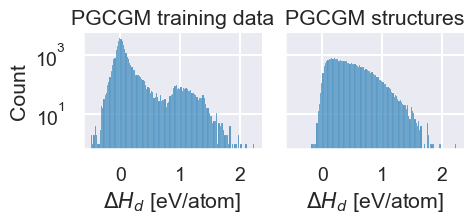}
    \caption{12,826 of the 33,172 \gls{PGCGM} training set structures are predicted to have a negative decomposition enthalpy (left) $\Delta H_d$ by \gls{ALIGNN}, indicating they are predicted to be thermodynamically stable. In contrast, all but 195 of the 27,116 structures generated by~\gls{PGCGM} were predicted to have a positive decomposition enthalpy (right). The median predicted decomposition enthalpy is also lower for the training data than the generated structures: 0.02 eV/atom vs. 0.38 eV/atom.}
    \label{fig:enthalpy_all}
\end{figure}

We evaluate \gls{ALIGNN} on \gls{PGCGM}'s training data ($33,172$ structures taken from \gls{MP}, \gls{ICSD}, and \gls{OQMD}) and the $27,116$ \gls{PGCGM}-generated structures, calculating both their predicted decomposition enthalpies $\Delta H_d$ and their embeddings $h$. In~\cref{fig:enthalpy_all}, we show that only $195$ of the $27,116$ generated structures are predicted by \gls{ALIGNN} to be thermodynamically stable. In contrast, $12,826$ of the $33,172$ training set structures are predicted to be stable.

\begin{figure}
    \centering
    \includegraphics[width=0.9\linewidth]{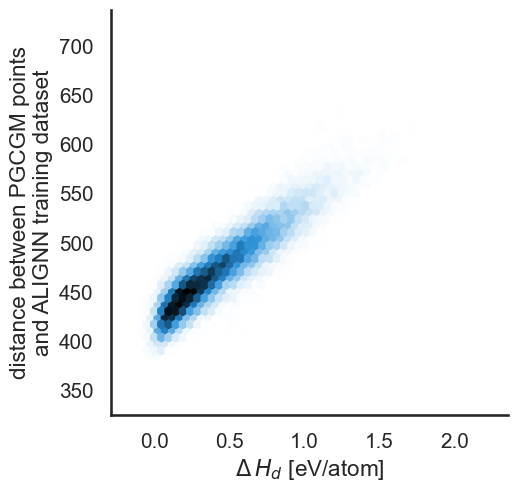}
    \caption{The minimum distance between each of the 27,117 generated materials' embeddings ($h$ in Eq.~\ref{eq:alignn}) and the \gls{ALIGNN} training data's embeddings has a strong positive correlation with the predicted decomposition enthalpy (shading indicates regions where points are concentrated). Thus, only the generated structures most similar to the training data are predicted to be thermodynamically stable.}
    \label{fig:stability_distance}
\end{figure}

We propose and evaluate two hypotheses for this result. First, the training data for and generated structures from \gls{PGCGM} are potentially out-of-domain for the trained \gls{ALIGNN} model. In~\cref{fig:stability_distance}, we show that there is a strong correlation between the minimum Euclidean distance of a \gls{PGCGM}-generated structure's embedding $h$ to the set of embeddings of the \gls{ALIGNN} training set and the predicted $\Delta H_d$. Thus, only the generated structures most similar to training data are predicted to be stable. In~\cref{fig:alignn_validation} in~\cref{sec:supplemental_figures}, we build on this distance vs. prediction for a held-out subset of the \gls{ALIGNN} training data; we show that the correlation of~\cref{fig:stability_distance} is not a general characteristic of \gls{ALIGNN} predictions.

% \alex{mathematical sin: ``compress'' can have a precise meaning that we're not using here}

\begin{figure}
    \centering
    \includegraphics[width=\linewidth]{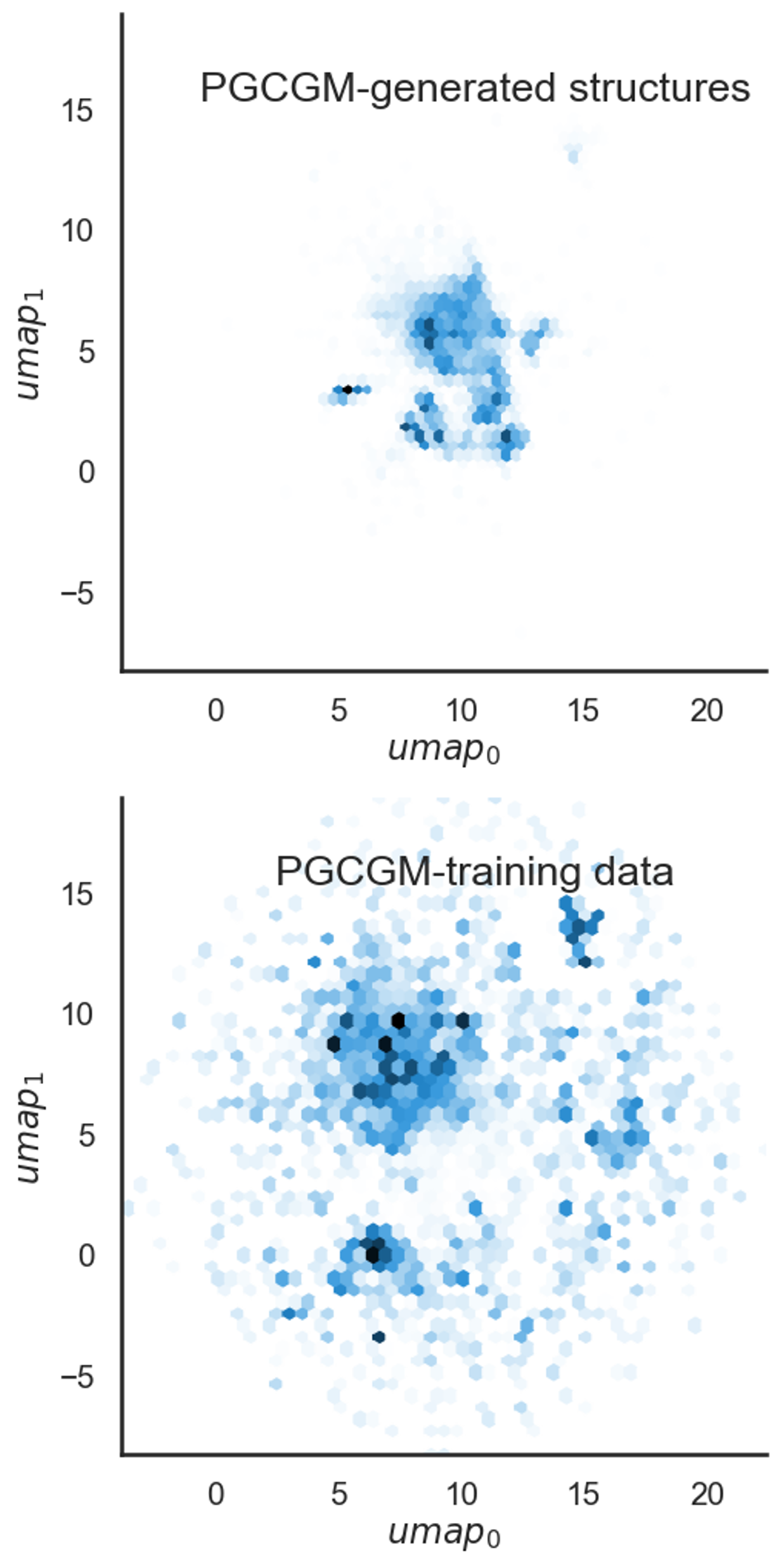}
    \caption{We use \gls{UMAP} to show that the \gls{ALIGNN} embeddings of the 27,116 valid \glspl{CIF} generated by \gls{PGCGM} are concentrated in a tight region of latent space. In contrast, the \gls{ALIGNN} embeddings of \gls{PGCGM}'s training data are spread more widely (shading indicates regions where points are concentrated). This suggests \gls{PGCGM}'s capacity for generating diverse and varied structures can be further improved. Here, the \gls{UMAP} embedding was learned from the combination of the \gls{PGCGM}'s generated structures and training data.}
    \label{fig:stability_umap}
\end{figure}

Beyond this distance-based analysis driving \gls{ALIGNN}'s predictions, we also provide evidence that the structures generated by \gls{PGCGM} are distributionally different and less varied than those in its training data. In~\cref{fig:stability_umap}, we take the embeddings of the \gls{PGCGM} training data and the generated structures and use \gls{UMAP}~\cite{Mcinnes2020umap} to project them into two dimensions. The embeddings of the generated structures are compressed into a tight region of the projection, while the embeddings of the training data are spread more widely across the space. Thus, there appears to be a lack of diversity in the generated structures, with respect to how the \gls{ALIGNN} model predicts their stability. This is similar to the general problem of mode collapse observed in \glspl{GAN}~\cite{Saxena2021modecollapse}, and strategies adopted there could mitigate it in future work.

In~\cref{fig:pgcgm_vs_training} (in the Appendix), we repeat this analysis using the embeddings of the \gls{PGCGM}-generated structures and \gls{ALIGNN}'s training data from \gls{MP}. As in~\cref{fig:stability_umap}, the embeddings of the generated structures are more compressed together than the embeddings of the training data.

% save out the embeddings ($h$ in Eq.~\ref{eq:alignn}) of its training data and our \gls{PGCGM}-generated structures. When projected into two dimensions using \gls{UMAP}~\cite{Mcinnes2020umap}, the embeddings of \gls{PGCGM}-generated structures are compressed into a much tighter region of the embedding space. This suggests a potential lack of diversity in the generated structures, with respect to \gls{PGCGM}'s training data. 

% \begin{figure*}
%     \centering
%     \includegraphics[width=0.75\linewidth]{figs/stability_umap_2.png}
%     \caption{The \gls{ALIGNN} embeddings of structures generated by \gls{PGCGM} are concentrated in a tight region of latent space, whereas the \gls{ALIGNN} embeddings of \gls{PGCGM}'s training data are spread more widely. This suggests \gls{PGCGM}'s capacity for generating diverse and varied structures can be further improved.}
%     \label{fig:stability_umap}
% \end{figure*}

\section{Conclusion}\label{sec:conclusion}

Here we assess the ability of generative structure models to assist in designing novel materials. Although current generative models like \gls{PGCGM} are able to generate large numbers of varied and valid material structures, we show that (1) the \gls{PGCGM} is not smooth with respect to either material validity or material property variation, making property optimization difficult; and (2) assessing thermodynamic stability of generated structures is difficult due to generated structures being concentrated in tight regions of latent space distant from property-predictor training data. Our particular analyses are not restricted to the \gls{PGCGM} and can be applied to assess other generative material models such as~\citet{Kim2022scgan,Ren2022ftcp,Court2020conddfcvae,Alverson2023discovery}.

% \christine{wondering if this paragraph can be rephrased as being related to future work? Maybe something like "Several potential investigations remain..."}
Our findings here suggest potential investigations to fully enable inverse design via generative modeling. For example, the loss formulation of \gls{PGCGM} or another generative model could be modified to encourage smoothness with respect to property variation in its input space via techniques like optimal transport~\cite{Chen2021optimal}. Global smoothness of \gls{PGCGM} input space is not necessarily needed, as functions that are locally Lipchitz can still be optimized in some settings~\cite{Patel2022lipschitz}.

Although we saw that most \gls{PGCGM}-generated structures are predicted to be unstable, this may be a consequence of a poor property-prediction model, not a poor input structure. Compared to general neural networks, graph neural networks like \gls{ALIGNN} have been shown to perform well in certain extrapolation tasks~\cite{Xu2021Domain}; however, the distance-to-training set metric we use in~\cref{fig:stability_distance} does not fully explain if a prediction is accurate~\cite{Hirschfeld2020uq}. Recent work also observes that \gls{ALIGNN} models can fail specifically in the setting of out-of-domain property prediction for material structures~\cite{Li2023outofdomain}. Challenges have also been identified in the use of graph neural networks to predict multiple material properties (e.g., stability and a design property of interest)~\cite{New2022curvatureinformed}. 

Here we analyze how diverse and stable structures directly predicted by the \gls{PGCGM} are. A supplementary strategy, taken by~\cite{Zhao2023pgcgm}, is to further refine \gls{ML}-predicted material structures, either with first-principles calculations, or with techniques like constrained Bayesian optimization~\cite{Zuo2021bowsr}.

\clearpage

\section*{Broader impact}

Inverse design of materials is important for many fields of engineering and science, and these fields have robust mechanisms for preventing misuse. Materials discoveries made by \gls{ML} can thus be beneficial to different aspects of society.
% \christine{maybe - *Material* discoveries made by...}

% Things to hit here:
% \begin{itemize}
%     \item Tying back to overall goal of inverse design
%     \item Potential pathologies in GNNs: curvature \cite{New2022curvatureinformed}, out-of-domain challenges: \cite{Xu2021Domain,Meredig2018LOCOCV}
%     \item Ways to make latent spaces more smooth~\cite{Chen2021optimal}
%     \item BOWSR-type methods to refine generated structures~\cite{Zuo2021bowsr}
%     \item \cite{Hirschfeld2020uq}
% \end{itemize}

\section*{Acknowledgements}
This work was supported by internal research and development funding from the Research and Exploratory Development Mission Area of the Johns Hopkins University Applied Physics Laboratory. We also thank the reviewers for their helpful suggestions and comments.

% \clearpage
\bibliography{references}
\bibliographystyle{synsml2023}

%%%%%%%%%%%%%%%%%%%%%%%%%%%%%%%%%%%%%%%%%%%%%%%%%%%%%%%%%%%%%%%%%%%%%%%%%%%%%%%
%%%%%%%%%%%%%%%%%%%%%%%%%%%%%%%%%%%%%%%%%%%%%%%%%%%%%%%%%%%%%%%%%%%%%%%%%%%%%%%
% APPENDIX
%%%%%%%%%%%%%%%%%%%%%%%%%%%%%%%%%%%%%%%%%%%%%%%%%%%%%%%%%%%%%%%%%%%%%%%%%%%%%%%
%%%%%%%%%%%%%%%%%%%%%%%%%%%%%%%%%%%%%%%%%%%%%%%%%%%%%%%%%%%%%%%%%%%%%%%%%%%%%%%
\newpage
\appendix
\onecolumn

% \section{Related work}\label{sec:related_work}

% \alex{with only 4 pages, we can probably skip any extended discussion of literature beyond what is strictly necessary.}

\section{Supplemental figures}\label{sec:supplemental_figures}

\begin{figure}[h]
    \centering
    \includegraphics[width=0.3\linewidth]{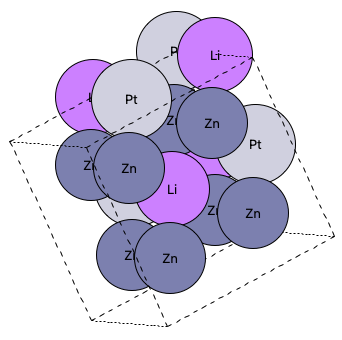}
    \caption{A \gls{PGCGM}-generated structure for $\mathrm{LiZn}_2\mathrm{Pt}$ with space group 225: $Fm\bar{3}m$ that is predicted by \gls{ALIGNN} to have a negative decomposition enthalpy $\Delta H_d$.}
    \label{fig:lizn2pt}
\end{figure}

\begin{figure}[h]
    \centering
    \includegraphics[width=0.3\linewidth]{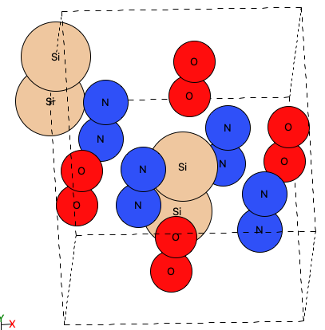}
    \caption{A \gls{PGCGM}-generated structure for $\mathrm{SiN}_2\mathrm{O}_2$ with space group 62: $Pnma$ predicted by \gls{ALIGNN} to have a positive decomposition enthalpy $\Delta H_d$.}
    \label{fig:sin2o2}
\end{figure}

% \begin{figure}
%     \centering
%     \includegraphics[width=\linewidth]{figs/generation_pvals.png}
%     \caption{With respect to the \gls{PGCGM}'s random input space ($z$ in Eq.~\ref{eq:generator}), for materials sampled from the $\mathrm{Ta{-}Ge{-}As}$ system with the $Pm{-}3m$ spacegroup, most components of $z$ do not have a significant Spearman correlation with whether or not $z$ is mapped to a valid \gls{CIF}, the resultant \gls{CIF}'s density, or the resultant \gls{CIF}'s unit cell volume. This lack of smoothness means that it will be difficult for gradient-based optimization to be used to improve properties of a material based on its input representation $z$.}
%     \label{fig:property_variation}
% \end{figure}

\begin{figure}[h]
    \centering
    \includegraphics[width=\linewidth]{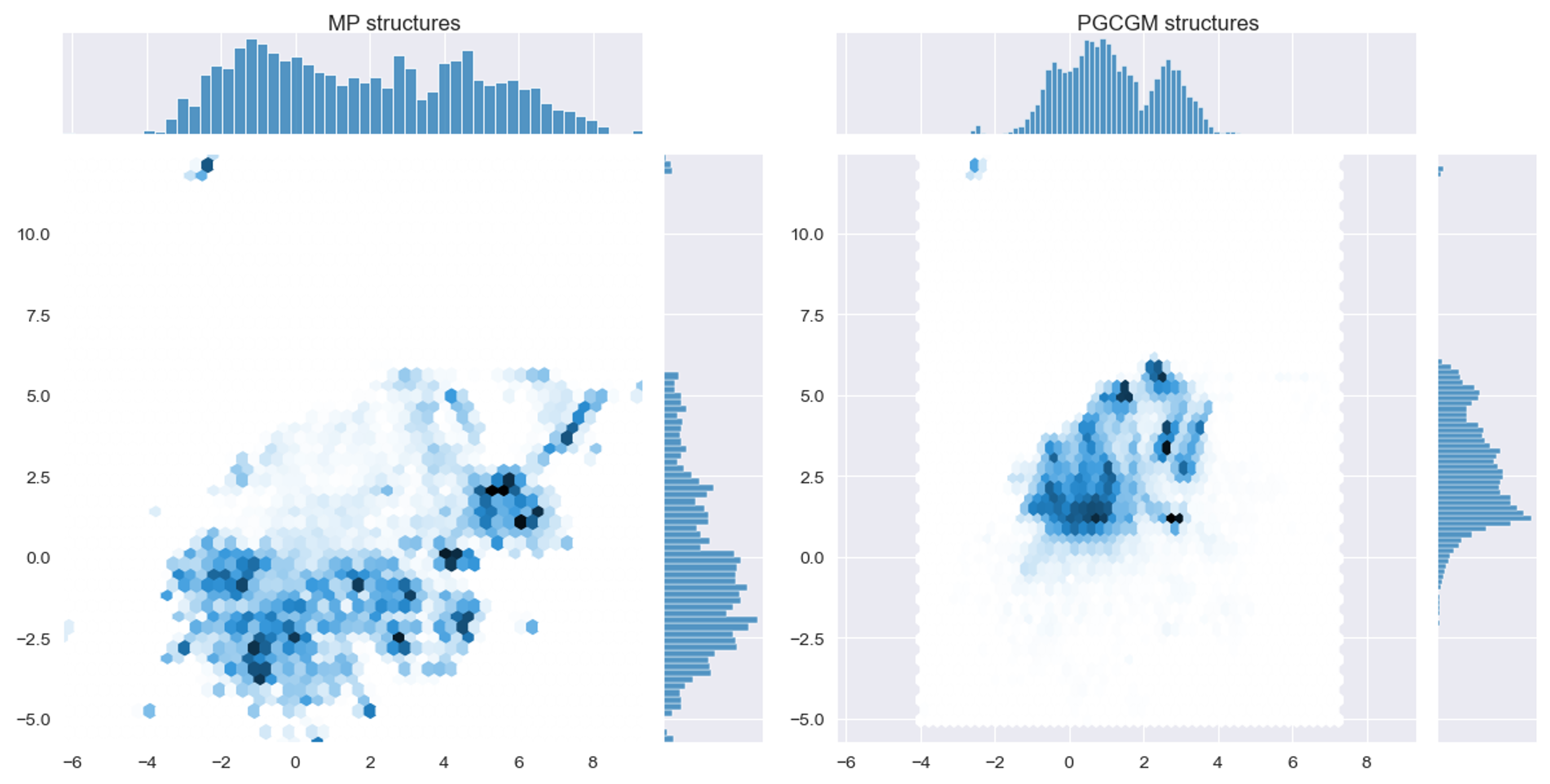}
    \caption{We use \gls{UMAP}~\cite{Mcinnes2020umap} and project the \gls{ALIGNN} embeddings of the 27,116 \gls{PGCGM}~\cite{Zhao2023pgcgm}-generated structures and the \gls{MP} structures of the \gls{ALIGNN} training data into two dimensions (shading indicates regions of high concentration). The \gls{PGCGM}-generated structures are much more compressed in the latent space than the embeddings of the \gls{ALIGNN} training data.}
    \label{fig:pgcgm_vs_training}
\end{figure}

\begin{figure}[h]
    \centering
    \includegraphics[width=0.5\linewidth]{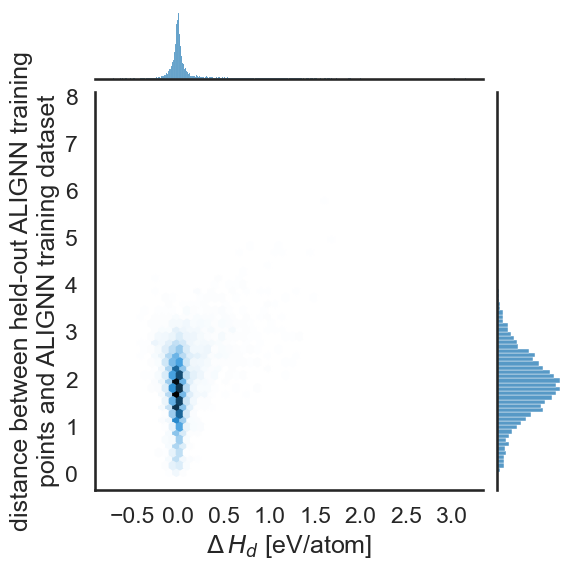}
    \caption{We repeat the distance-to-training-set analysis of~\cref{fig:stability_distance} while only using~\gls{ALIGNN}'s~\cite{Choudhary2021alignn} training data. We hold out 20\% ($7,568$ structures) of its training dataset, and we calculate the minimal distance from each point to the remaining $30,274$ training points. Unlike~\cref{fig:stability_distance}, there is no clear correlation between the \gls{ALIGNN}'s predicted decomposition enthalpy and the distance to the remaining training points. This suggests that the correlation between predicted decomposition enthalpy and distance-to-training set occurs specifically in the case of out-of-domain data.}
    \label{fig:alignn_validation}
\end{figure}

\clearpage

\section{Data details}\label{sec:data}

For \gls{PGCGM}~\cite{Zhao2023pgcgm}, we used the pre-trained model, which was trained on 33,172 ternary structures sourced from \gls{MP}~\cite{Jain2013mp}, \gls{ICSD}~\cite{Belsky2002icsd}, and \gls{OQMD}~\cite{Saal2013OQMD}. The IDs were made available by the \gls{PGCGM} authors on GitHub\footnote{\url{https://github.com/MilesZhao/PGCGM/blob/main/data/ids_for_mp_oqmd_icsd.csv}}. We trained our \gls{ALIGNN}~\cite{Choudhary2021alignn} model on the 37,842 ternary structures from \gls{MP}, selected from the 85,014 structures analyzed by \citealt{Bartel2020stability}.

\end{document}

%% file: glossary.tex
% Project-Scope Glossary
% A ---------------------------------------------------------------------------
\newacronym{APL}{APL}{%
  the Johns Hopkins University Applied Physics Laboratory%
}
\newacronym{ALIGNN}{ALIGNN}{atomistic line graph neural network}

% B ---------------------------------------------------------------------------

% C ---------------------------------------------------------------------------
\newacronym{CIF}{CIF}{crystallographic information file}

% D ---------------------------------------------------------------------------

% E ---------------------------------------------------------------------------

% F ---------------------------------------------------------------------------
\newacronym{FTCP}{FTCP}{Fourier-transformed crystal properties}

% G ---------------------------------------------------------------------------
\newacronym{GAN}{GAN}{generative adversarial network}
\newacronym{GNN}{GNN}{graph neural network}

% H ---------------------------------------------------------------------------

% I ---------------------------------------------------------------------------
\newacronym{ICSD}{ICSD}{Inorganic Crystal Structure Database}

% J ---------------------------------------------------------------------------
\newacronym{JHU}{JHU}{%
  Johns Hopkins University%
}

% K ---------------------------------------------------------------------------
\newacronym{KS}{KS}{Kolmorogov-Smirnov}

% L ---------------------------------------------------------------------------

% M ---------------------------------------------------------------------------
\newacronym{ML}{ML}{machine learning}
\newacronym{MLP}{MLP}{multi-layer perceptron}
\newacronym{MP}{MP}{Materials Project}

% N ---------------------------------------------------------------------------

% O ---------------------------------------------------------------------------
\newacronym{OQMD}{OQMD}{Open Quantum Materials Database}

% P ---------------------------------------------------------------------------
\newacronym{PGCGM}{PGCGM}{physics-guided crystal generation model}
\newacronym{PCA}{PCA}{principal component analysis}

% R ---------------------------------------------------------------------------

% S ---------------------------------------------------------------------------
% T ---------------------------------------------------------------------------
% U ---------------------------------------------------------------------------
\newacronym{UMAP}{UMAP}{uniform manifold approximation and projection}

% V ---------------------------------------------------------------------------
\newacronym{VAE}{VAE}{variational autoencoder}

% X ---------------------------------------------------------------------------